\documentclass{PoS}

\def\ie{{i.e.,~}}
\def\lat{{{\it Fermi}-LAT~}}

\title{Extragalactic source population studies at very high energies in the Cherenkov Telescope Array era}

\ShortTitle{Extragalactic source population studies at VHE in the CTA era}
        
\author{\speaker{T. Hassan}$^{1}$, A. Dom{\'i}nguez$^{2}$, J. Lefaucheur$^{3}$, D. Mazin$^{4}$, S. Pita$^{5}$, A. Zech$^{3}$ for the CTA Consortium\\
       $^{1}$Institut de Fisica d'Altes Energies (IFAE), The Barcelona Institute of Science and Technology\\
       $^{2}$Universidad Complutense de Madrid, Grupo de Altas Energ\'{i}as (GAE)\\
       $^{3}$LUTH, Observatoire de Paris, PSL Research University, CNRS\\
       $^{4}$University of Tokyo, ICRR\\
       $^{5}$APC, AstroParticule et Cosmologie, Universit\'e Paris Diderot, CNRS/IN2P3, CEA/Irfu, Observatoire de Paris\\
       E-mail: \email{thassan@ifae.es, alberto@gae.ucm.es}}

\abstract{The Cherenkov Telescope Array (CTA) is the next generation ground-based $\gamma$-ray observatory. It will provide an order of magnitude better sensitivity and an extended energy coverage, 20 GeV -- 300 TeV, relative to current Imaging Atmospheric Cherenkov Telescopes (IACTs). IACTs, despite featuring an excellent sensitivity, are characterized by a limited field of view that makes the blind search of new sources very time inefficient. Fortunately, the {\it Fermi}-LAT collaboration recently released a new catalog of 1,556 sources detected in the 10 GeV -- 2 TeV range by the Large Area Telescope (LAT) in the first 7 years of its operation (the 3FHL catalog). This catalog is currently the most appropriate description of the sky that will be energetically accessible to CTA. Here, we discuss a detailed analysis of the extragalactic source population (mostly blazars) that will be studied in the near future by CTA. This analysis is based on simulations built from the expected array configurations and information reported in the 3FHL catalog. These results show the improvements that CTA will provide on the extragalactic TeV source population studies, which will be carried out by Key Science Projects as well as dedicated proposals.

}

\FullConference{35th International Cosmic Ray Conference - ICRC2017\\
		10-20 July, 2017\\
		Bexco, Busan, Korea}

\begin{document}

\section{Introduction}
\label{sec:intro}

During the last decade, the success of the {\it Fermi}-Large Area Telescope (LAT) and the development and refinement of the Atmospheric Cherenkov Telescope (IACT) technique by MAGIC, H.E.S.S., and VERITAS has revolutionised our understanding of the non-thermal high-energy Universe. The next generation of IACTs is currently under development, led by the Cherenkov Telescope Array (CTA).\footnote{http://www.cta-observatory.org/} CTA will provide unprecedented insights into the $\gamma$-ray sky from 20 GeV to 300 TeV, improving the sensitivity of current IACTs by more than an order of magnitude.

CTA will be composed of two observatories providing (near to) full-sky coverage. One array will be located in the Northern Hemisphere at La Palma (Spain) and a second array in the Southern Hemisphere at Paranal (Chile). To maximise the scientific output, these arrays are planned to have different designs: the Northern array (hereafter, CTA-N) is planned to be composed by 4 Large-Sized Telescopes (LST) and 15 Medium-Sized Telescopes (MST), whereas the Southern array (hereafter, CTA-S) is planned to be larger, composed by 4 LSTs, 25 MSTs, and 70 Small-Sized Telescopes (SST). 

The combination of space-borne $\gamma$-ray telescopes and ground-based IACTs has proven to be successful on expanding our knowledge on persistent as well as transient phenomena. First, the large collection area of IACTs makes up for the limited payload of space telescopes, allowing us to explore shorter variability time scales with an improved sensitivity for moderate observation times \cite{fermi_vs_CTA}. Second, space-borne instruments, in particular the {\it Fermi}-LAT, compensates for the limited field of view of IACTs, typically of a few degrees in diameter, as it surveys the whole sky in a matter of hours, having accumulated observation time over the whole $\gamma$-ray sky for nearly a decade. For this reason, the all-sky survey conducted by the \lat has been a key asset for ground-based telescopes, guiding follow-up observations on potential very-high-energy ($E>100$~GeV) emitters and triggering targets of opportunity on transient phenomena.

The CTA improved differential sensitivity will enable the observation of fainter sources, significantly increasing the detectable population of Active Galactic Nuclei (AGN), including their average (quiescent) flux states, which in general are not detectable by the current generation of IACTs. Here, we use the information provided above 10~GeV by the Third Catalog of Hard \lat Sources (the 3FHL catalog, \cite{3FHL}) to make predictions on the extragalactic source populations that both CTA-N and CTA-S will be able to detect over short and moderate telescope exposures. Note that the results presented here make use of average flux states, \ie integrated over the 7 years of LAT exposure, and do not take into account the strong variability of these sources.

\section{Description of the 3FHL catalog}
\label{sec:catalog}
The \lat surveys the whole sky every three hours with excellent sensitivity and angular resolution. The LAT Collaboration recently released the 3FHL catalog, which describes the hardest $\gamma$-ray sources in the sky by increasing the lower energy threshold of the analysis to 10~GeV \cite{3FHL}, relative to the broad-band catalog 3FGL \cite{3FGL}. The 3FHL is built from 7 years of Pass~8 data (while the 3FGL contained 4 years), which provides several improvements in comparison with previous versions of the event-level analysis. The 3FHL lists the position and spectral characteristics of 1,556 sources over the whole sky. Most of these sources 1,231 (79\% of the total catalog) are associated with sources of extragalactic nature, and 526 (43\% of the extragalactic sources) have a known redshift. Note that only 72 of the 3FHL extragalactic sources have been already detected by ground-based telescopes.\footnote{See: http://tevcat.uchicago.edu/}

Given the low energy threshold of the future CTA, expected to be of approximately 20~GeV, the 3FHL is the best available sample of targets for the observatory, providing an excellent opportunity to derive robust and realistic predictions on the persistent sources that will be accessible in the near future.

\section{CTA detectability computation}
\label{sec:simulations}

The 3FHL provides spectral fluxes in 5 energy bands from 10~GeV to 2~TeV. The energy limits of these bands are 10, 20, 50, 150, 500 GeV and 2 TeV. For most cases, only upper limits are given for the higher energy bands. Since CTA will be sensitive to photons with energies $>1$~TeV, there are several steps that need to be performed with caution to extrapolate \lat blazar spectra to higher energies:

\begin{itemize}    
    \item \textbf{Spectral shape of the intrinsic emission:} Given the 3FHL energy range and the low photon statistics at TeV energies, the spectral flux extrapolation to higher energies is extremely dependent on the function considered to fit the \lat fluxes.
    
    \item \textbf{Extragalactic background light (EBL):} We must take into account the flux attenuation produced by the pair-production interaction between $\gamma$-rays traveling over cosmological distances and photons from the diffuse extragalactic background light (EBL), e.g. \cite{dominguez11a}. This flux attenuation depends on the energy of the $\gamma$-ray photon and distance to the source.
    
\end{itemize}

We approach these issues as follows. First, by testing different flux extrapolations to the TeV range. These extrapolations are modeled with the following functions that include the exponential attenuation from the EBL effect. Second, assuming the flux attenuations provided by \cite{dominguez11a}, which are compatible with the current EBL knowledge.

\begin{itemize}
    \item Power-law + EBL attenuation (PL)
    \item Power-law with exponential cutoff + EBL attenuation. An exponential cutoff is added to the power-law at $1/(1+z)$ TeV (PLE)
    \item Broken Power-law + EBL attenuation. For hard sources (with a power-law index $\Gamma$ > 2), spectra are softened to $\Gamma = 2.5$ at $100/(1+z)$ GeV (BPL)
    \item Log-Parabola + EBL attenuation (LP)
\end{itemize}

Note that we are listing above the extrapolation scenarios from most optimistic to most pessimistic, meaning that a PL will predict a larger flux for a given TeV energy than a LP.

\begin{figure*}
\begin{center}
\includegraphics[width=7cm]{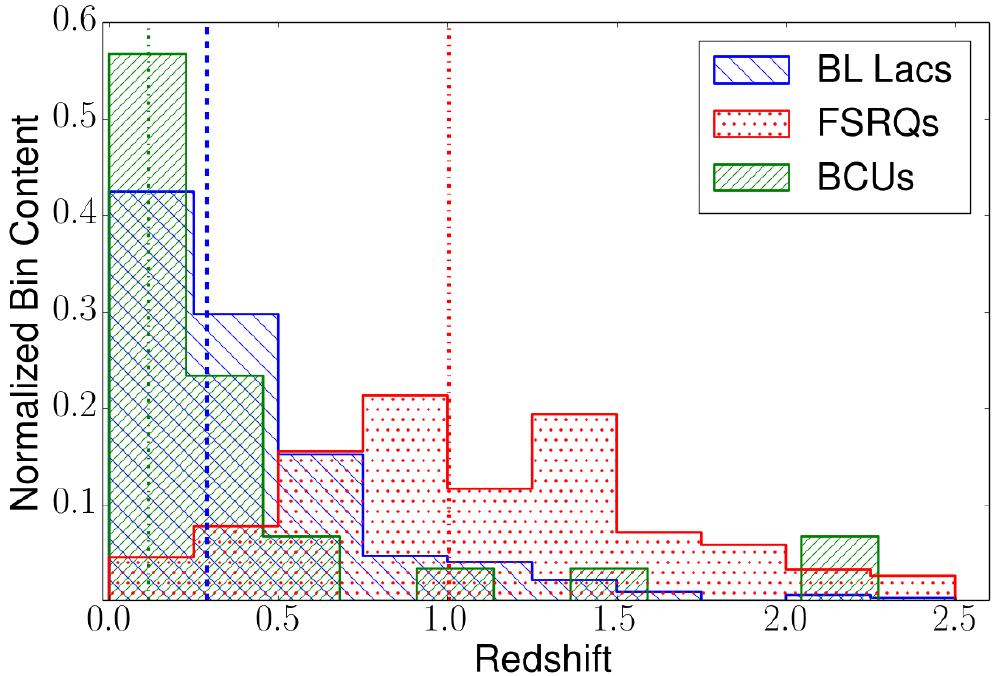}\\
\caption{Distribution of known redshifts of 3FHL extragalactic sources divided by their class. The median redshifts are 0.29, 1.00 and 0.12 for BL Lacs (blue), FSRQs (red), and BCUs (green) respectively. These values are plotted with vertical lines.}
\label{fig:redshift_3fhl_types}
\end{center}
\end{figure*}

As mentioned in Section \ref{sec:catalog}, a large fraction ($\sim 57$\%) of the 3FHL extragalactic sources do not have a known redshift. This situation adds an additional uncertainty to our flux extrapolations. We address this problem in this way. The 3FHL provides source classes for most of the extragalactic objects. The redshift distributions for BL Lacs, flat-spectrum radio quasars (FSRQs), and blazars of uncertain type (BCUs) are shown in Figure~\ref{fig:redshift_3fhl_types}. We sample randomly these distributions, according to their source class, to attach a redshift to each source of unknown redshift. This procedure is of course not expected to give robust redshifts for individual blazars but should work in a statistical sense, which is the goal of our analysis. Note that we do not take into account the possible bias induced by the increased difficulty in measuring the redshift of a source with increasing distance, which would be difficult to model.

Several tools are used for the spectral simulations and detectability forecasts, obtaining consistent results between them. These tools are CTAmacros,\footnote{See https://github.com/cta-observatory/ctamacros} \textit{GAEtools} \cite{GAEtools} and Gammapy \cite{gammaPy}. They use instrument response functions (IRFs) generated from detailed Monte Carlo simulations, which allow a CTA performance estimate (these IRFs correspond to the third large-scale production). Common conditions are used in these tools to claim a detection: 5$\sigma$ significance level assuming an off-to-on source exposure ratio of 5, with a minimum number of 10 excess events, and, at least, five times the expected systematic uncertainty in the background estimation, which is about 1\%.

As described in \cite{gernot, CTA_MC}, these IRFs were generated using analysis cuts that optimize differential sensitivities for point-like sources, which is ideal for our science case. At the moment, given the huge computation resources required for the production of these IRFs, they have been calculated only for zenith angles of 20$^\circ$ and 40$^\circ$. In our analysis, we estimate the altitude of culmination of each simulated source from each site (La Palma and Paranal, with latitudes 28.76$^\circ$ N and 24.63$^\circ$ S, respectively). For sources culminating between 0 and 30$^\circ$ in zenith angle, the IRF corresponding to 20$^\circ$ is used, whereas sources that culminate at zenith angles of 30 to 50$^\circ$ are simulated using the 40$^\circ$ IRF. We assume that no source is detectable if it culminates at zenith angles larger than 50$^\circ$. This condition excludes only less than 5\% of the extragalactic sample from both of the CTA sites together.

\section{Results}
\label{sec:results}

Table~\ref{table:number_of_sources} shows the number of extragalactic sources detected for each of the proposed spectral models along with the number of those sources already detected with current IACTs. This TeV information is provided by the 3FHL catalog. These calculations follow the recipe detailed in Section \ref{sec:simulations}. As expected, the total number of detectable sources is strongly affected by the intrinsic emission model that is assumed. 

\begin{table}[htp]

\begin{center}
\begin{tabular}{l|c|c|c|c||c|c}
 & CTA-N & CTA-N & CTA-S & CTA-S & CTA-N/S & CTA-N/S\\
 & (5h) & (20h) & (5h) & (20h) & (5h) & (20h)\\
\hline

PL  & 261 (50)  & 383 (54) & 272 (37) & 381 (40) & 395 (63) & 566 (69) \\
PLE & 190 (47)  & 339 (52) & 190 (37) & 336 (39) & 304 (62) & 498 (66) \\
BPL & 142 (45)  & 302 (53) & 142 (45) & 322 (39) & 238 (57) & 470 (68) \\
LP  & 67  (37)  & 139 (47) & 75  (27) & 161 (37) & 111 (50) & 230 (62) \\

\end{tabular}
\end{center}
\caption{\label{table:number_of_sources} Number of 3FHL extragalactic sources detectable by CTA under different extrapolation schemes and two different observation times. From each sub-sample of detectable sources, the number of these objects already detected by the current generation of IACTs is shown in parenthesis. First four columns refer to the number of detected sources observable by CTA-N and CTA-S independently, while the last two columns show the total number of sources detectable by any of them (shown in Fig. \ref{fig:allsky}, for PLE extrapolation).}
\end{table}

Note the LP fits should be taken with caution as existing TeV observations disfavour such an extrapolation of LAT spectra \cite{gpropa}. In the 3FHL, for a typical blazar, there are detections only at the lower energy bins, \ie approximately between 10 and 150~GeV, therefore a log-parabola extrapolation tends to predict TeV fluxes with a significantly stronger curvature than the one observed by IACTs. Complementarily, \cite{gpropa} discusses that using a PLE extrapolation roughly reproduces average flux states observed at TeV energies. Under this extrapolation scheme, the number of detected sources as a function of their class is shown in Table~\ref{table:sources_per_class}.

\begin{table}[htp]
\begin{center}
\begin{tabular}{l|c|c|c|c||c|c}
 & CTA-N & CTA-N & CTA-S & CTA-S & CTA-N/S & CTA-N/S\\
 & [5h] & [20h] & [5h] & [20h] & [5h] & [20h]\\
\hline

BL Lacs  			& 136 (39) & 243 (43) & 138 (27) & 223 (30) & 208 (50) & 344 (53) \\
FSRQs    			& 6 (3)    & 9 (4)   & 6 (3)     & 10 (3)   & 7 (3)    & 13 (4) \\
Blazar of uncertain type     	& 43 (2)   & 80 (2)   & 62 (2)   & 94 (2)   & 79 (3)   & 129 (3) \\
Radio galaxy     		& 4 (3)    & 6 (3)    & 5 (3)    & 6 (3)    & 7 (5)    & 9 (5) \\
Non-blazar active galaxy     	& 1 (0)    & 1 (0)    & 1 (0)    & 1 (0)    & 1 (0)    & 1 (0) \\
Star Burst Galaxies  		& 0 (0)    & 0 (0)    & 2 (1)    & 2 (1)    & 2 (1)    & 2 (1) \\

\end{tabular}
\end{center}
\caption{\label{table:sources_per_class} Number of 3FHL extragalactic sources detectable by CTA divided by their class, under the PLE extrapolation scheme, for two different observation times. From each sub-sample of detectable sources, the number of these objects already detected by the current generation of IACTs is shown in parenthesis.}
\end{table}

\begin{figure*}
\begin{center}
\includegraphics[width=7.5cm]{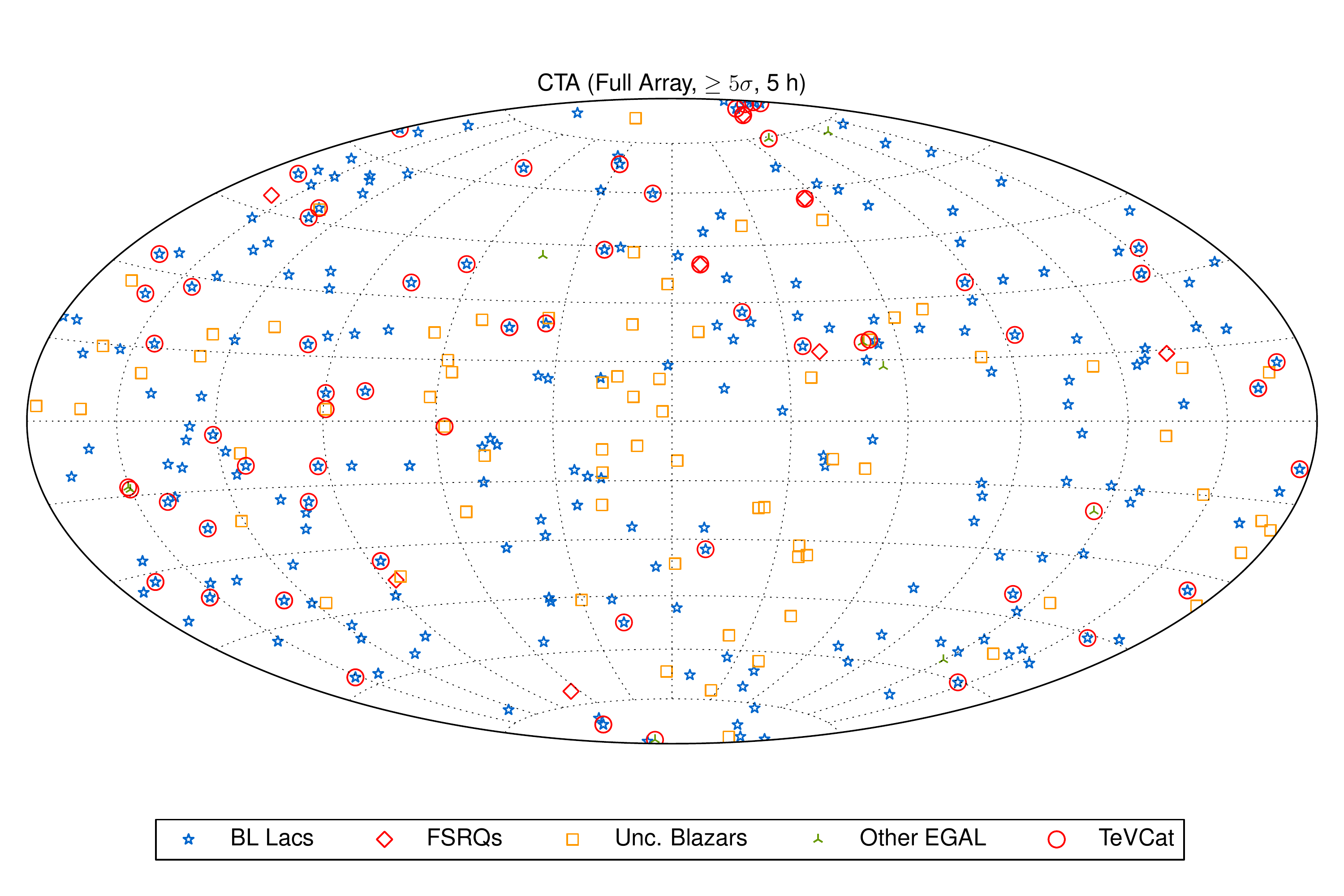}
\includegraphics[width=7.5cm]{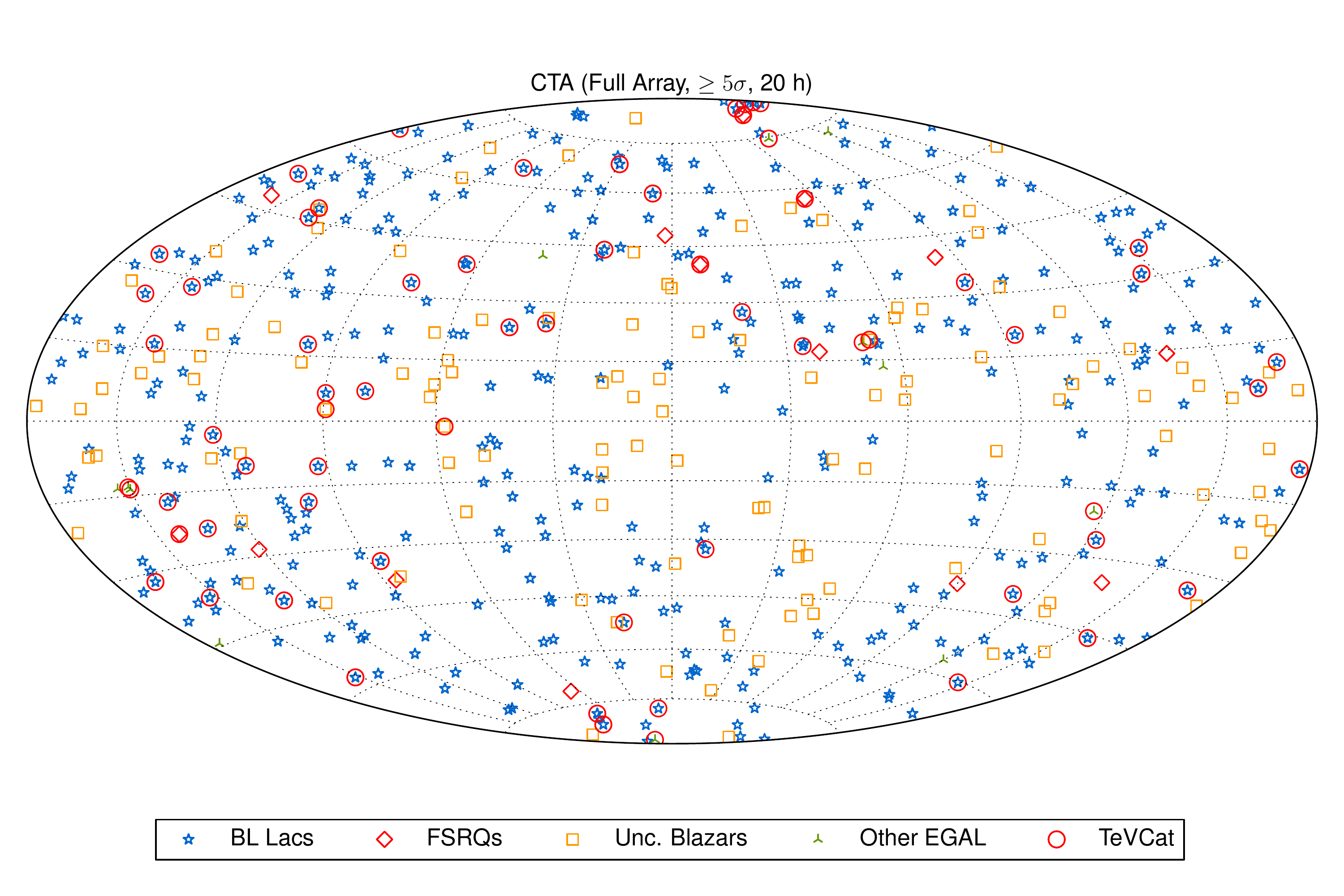}
\caption{Skymaps of the predicted extragalactic sources detectable by the CTA Observatory (either CTA-N or CTA-S), under the PLE extrapolation scheme, at $S \geq 5\sigma$ in 5~h (\emph{left}) and 20~h (\emph{right}). Different symbols and colours are used for different source populations. Galactic coordinates and Hammer-Aitoff projection are used. Note the population of sources with unknown redshift (more than 50\% of the total sample) are assumed to have a random redshift, following the same distribution as the ones with known redshift.}
\label{fig:allsky}
\end{center}
\end{figure*}

Figure~\ref{fig:allsky} shows in Galactic coordinates (Hammer-Aitoff projection) all extragalactic sources that are predicted to be detected ($\geq 5 \sigma$) by CTA-N and CTA-S assuming the PLE extrapolations. This figure shows the improvements that CTA will provide on the extragalactic TeV source population studies, which will be carried out by Key Science Projects as well as dedicated proposals.

\begin{figure*}
\includegraphics[width=7cm]{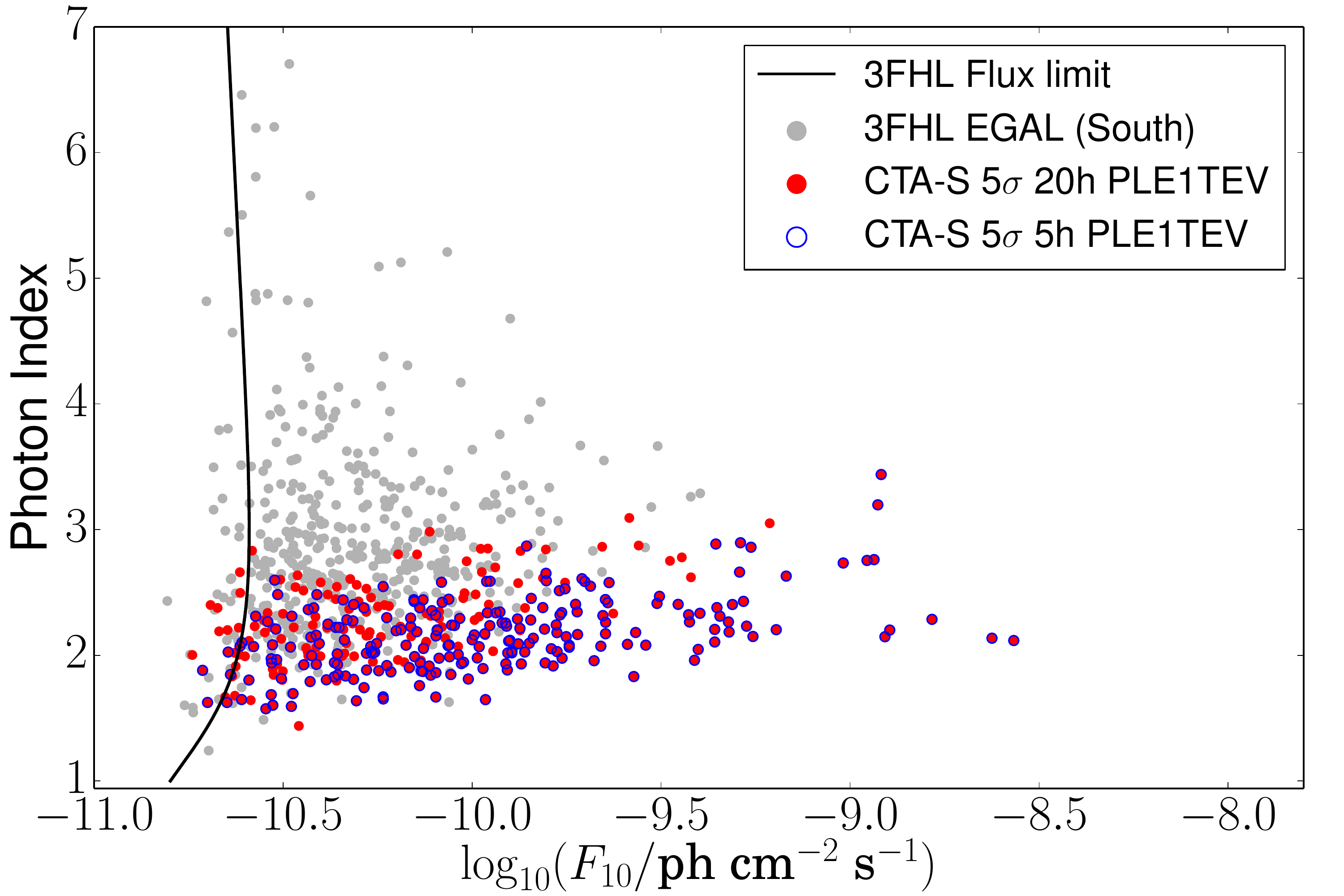}
\includegraphics[width=7cm]{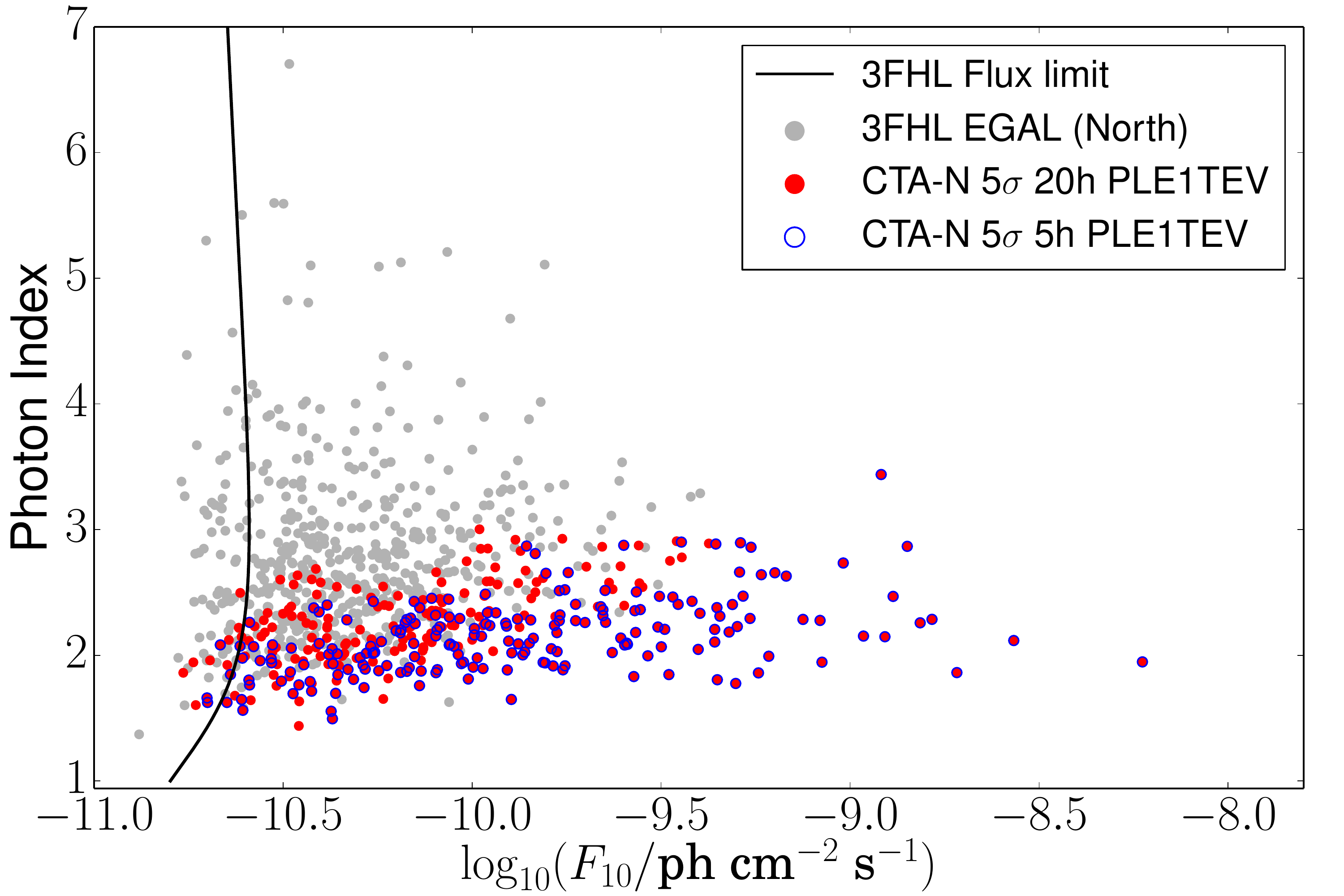}\\
\caption{(\textit{Left}) Photon index as a function of integrated flux above 10 GeV in the LAT energy band. Assuming a PLE extrapolation, we show 3FHL sources that are visible from CTA-S at culminations lower than $50^{\circ}$ (gray circles), 3FHL sources detectable in 20h of observations (red filled circles), and soures detectable in 5h (blue empty circles). The 3FHL flux limit is given by \cite{3FHL}. (\textit{Right}) Same for CTA-N.}
\label{fig:index_vs_flux}
\end{figure*}

Figure~\ref{fig:index_vs_flux} shows the photon index versus integrated flux (in the LAT energy band) above 10 GeV of all 3FHL extragalactic sources that are visible either from CTA-N or CTA-S (at culminations lower than $50^{\circ}$), in comparison with those that could be detected by CTA in 5 and 20h of telescope exposure. Interestingly, we can see in Figure~\ref{fig:index_vs_flux} that even in 5h, there are sources detected at the 3FHL flux limit. This result indicates that CTA could probably detect a new population of sources with hard spectra and low fluxes still not seen by the LAT.

\begin{figure*}
\includegraphics[width=7cm]{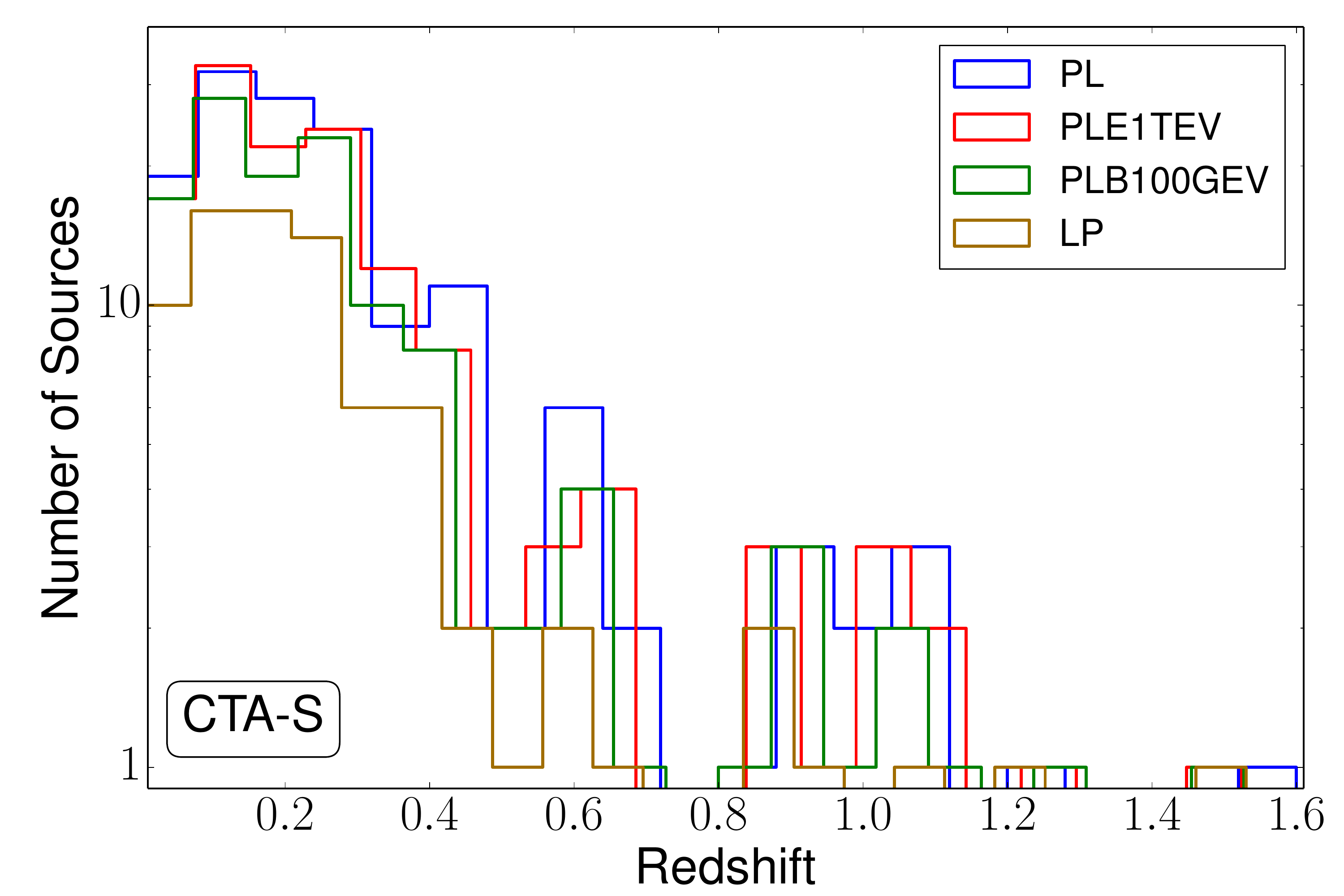}
\includegraphics[width=7cm]{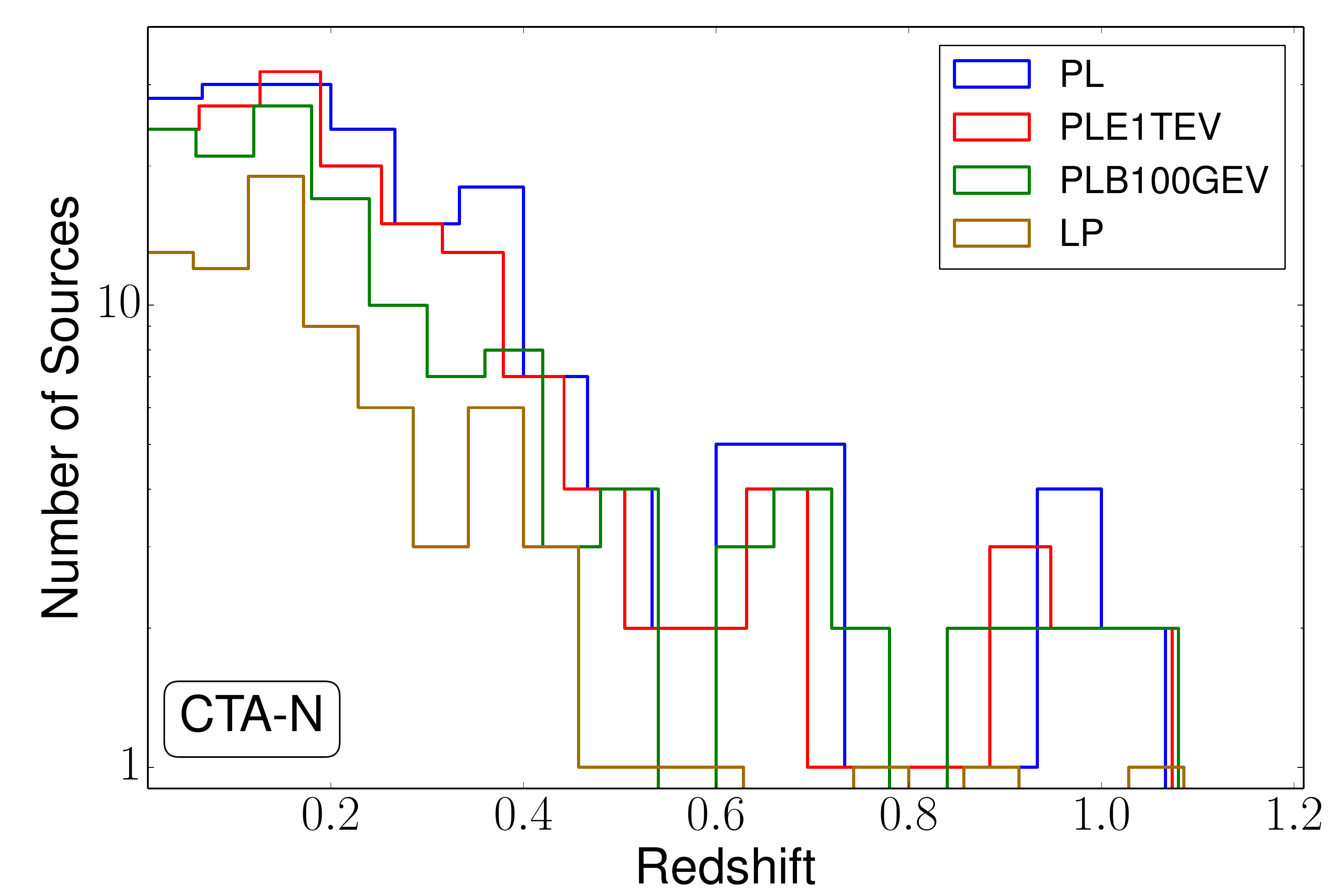}
\caption{(\emph{Left panel}) Number of sources detected by CTA-S as a function of redshift for different flux extrapolations (see text for details) (\emph{Right panel}) Same for CTA-N. Note the highest redshift detectable blazars (within the 1 -- 1.5 range) are of known redshift.}
\label{fig:number_vs_redshift}
\end{figure*}

CTA will not just increase the number of detected extragalactic sources in the TeV regime, it will also expand the frontier on the farthest VHE sources detected from the ground. As shown in Figure~\ref{fig:number_vs_redshift}, a significant number of high redshift sources in their average state will be detectable by CTA, a few of them even as far as $z\sim 1.5$. These blazars at high redshift are appealing for different scientific topics, such as estimating the EBL spectral intensities and their evolution \cite{CTA_EBL,dominguez13a,dominguez15}, studying intergalactic magnetic fields \cite{gpropa} and axion-like particles \cite{dominguez11b}, evaluating cosmological properties \cite{dominguez13b} and also testing Lorentz invariance violations \cite{CTA_LIV}. It is well known that blazars suffer flaring episodes that will make their detectability possible in shorter exposures, and will enable the detection of sources at even higher redshifts \cite{B0218, PKS1441,1441_veritas}. 

\section{Conclusions}
\label{sec:conclusions}

These results show the CTA potential to expand our understanding of the extragalactic populations of $\gamma$-ray emitters and their evolution over redshift. CTA will not only dramatically increase the amount of detectable blazars in their average state, it will also enable their spectral study over more than five decades in energy by combining its observations with \lat data. Above 10~GeV the sky is completely dominated by BL Lacs \cite{3FHL}, however we have shown that with moderate observation times, there will be other classes of extragalactic source detected in their average state such as tens of FSRQs, radio galaxies, and starburst galaxies. We have also shown the CTA potential in discovering new source populations of hard spectra and low fluxes. As mentioned above, there are fundamental science topics that will benefit from the large number of extragalactic sources that CTA will detect, such as questions related with gamma-ray propagation on cosmic scales.

We stress that our results are based on the more persistent sky and that we should expect a large amount of detections of new flaring sources, which are not included in our analysis. Information provided by the \lat will be key to optimising follow-up observations strategies for CTA. An important uncertainty in using our spectral flux extrapolations comes from the large number of \lat blazars with unknown redshift. We encourage dedicated observational campaigns to reduce these uncertainties and also guide follow-up observations \cite{redshifts}.

\subsubsection*{Acknowledgments}
This work was conducted in the context of the CTA Extragalactic Working Group. We gratefully acknowledge financial support from the agencies and organizations listed here : http://www.cta-observatory.org/consortium\_acknowledgments

\bibliographystyle{JHEP}
\bibliography{references}

\providecommand{\href}[2]{#2}\begingroup\raggedright\begin{thebibliography}{10}

\bibitem{fermi_vs_CTA}
S.~{Funk}, J.~A. {Hinton} and {CTA Consortium}, \emph{{Comparison of Fermi-LAT
  and CTA in the region between 10-100 GeV}},
  \href{http://dx.doi.org/10.1016/j.astropartphys.2012.05.018}{\emph{APP}
  {\bfseries 43} (Mar., 2013) 348--355},
  [\href{https://arxiv.org/abs/1205.0832}{{\ttfamily 1205.0832}}].

\bibitem{3FHL}
M.~{Ajelo at al.}, \emph{{3FHL: The Third Catalog of Hard Fermi-LAT Sources}},
  {\emph{ArXiv e-prints} (Feb., 2017) },
  [\href{https://arxiv.org/abs/1702.00664}{{\ttfamily 1702.00664}}].

\bibitem{3FGL}
F.~{Acero} and {Fermi-LAT Collaboration}, \emph{{Fermi Large Area Telescope
  Third Source Catalog}},
  \href{http://dx.doi.org/10.1088/0067-0049/218/2/23}{\emph{APJS} {\bfseries
  218} (June, 2015) 23}, [\href{https://arxiv.org/abs/1501.02003}{{\ttfamily
  1501.02003}}].

\bibitem{dominguez11a}
A.~{Dom{\'{\i}}nguez} et~al., \emph{{Extragalactic background light inferred
  from AEGIS galaxy-SED-type fractions}},
  \href{http://dx.doi.org/10.1111/j.1365-2966.2010.17631.x}{\emph{MNRAS}
  {\bfseries 410} (Feb., 2011) 2556--2578},
  [\href{https://arxiv.org/abs/1007.1459}{{\ttfamily 1007.1459}}].

\bibitem{GAEtools}
T.~{Hassan}, \emph{{Sensitivity studies for the Cherenkov Telescope Array}},
  {PhD} dissertation, Universidad Complutense de Madrid, Atomic and Nuclear
  Physics Department, 2015.

\bibitem{gammaPy}
A.~{Donath} et~al., \emph{{Gammapy: An open-source Python package for gamma-ray
  astronomy}},  in \emph{34th ICRC}, vol.~34 of \emph{International Cosmic Ray
  Conference}, p.~789, July, 2015,
  \href{https://arxiv.org/abs/1509.07408}{{\ttfamily 1509.07408}}.

\bibitem{gernot}
G.~{Maier at al. for the CTA Consortium}, \emph{{Performance of the Cherenkov
  Telescope Array}}, .

\bibitem{CTA_MC}
T.~{Hassan} et~al., \emph{{Monte Carlo Performance Studies for the Site
  Selection of the Cherenkov Telescope Array}}, {\emph{ArXiv e-prints} (May,
  2017) }, [\href{https://arxiv.org/abs/1705.01790}{{\ttfamily 1705.01790}}].

\bibitem{gpropa}
F.~{Gate et al. for the CTA Consortium}, \emph{{Studying cosmological gamma-ray
  propagation with the Cherenkov Telescope Array}}, .

\bibitem{CTA_EBL}
D.~{Mazin}, others and {CTA Consortium}, \emph{{Potential of EBL and cosmology
  studies with the Cherenkov Telescope Array}},
  \href{http://dx.doi.org/10.1016/j.astropartphys.2012.09.002}{\emph{APP}
  {\bfseries 43} (Mar., 2013) 241--251},
  [\href{https://arxiv.org/abs/1303.7124}{{\ttfamily 1303.7124}}].

\bibitem{dominguez13a}
A.~{Dom{\'{\i}}nguez} et~al., \emph{{Detection of the Cosmic {$\gamma$}-Ray
  Horizon from Multiwavelength Observations of Blazars}},
  \href{http://dx.doi.org/10.1088/0004-637X/770/1/77}{\emph{APJ} {\bfseries
  770} (June, 2013) 77}, [\href{https://arxiv.org/abs/1305.2162}{{\ttfamily
  1305.2162}}].

\bibitem{dominguez15}
A.~{Dom{\'{\i}}nguez} and M.~{Ajello}, \emph{{Spectral Analysis of Fermi-LAT
  Blazars above 50 GeV}},
  \href{http://dx.doi.org/10.1088/2041-8205/813/2/L34}{\emph{APJl} {\bfseries
  813} (Nov., 2015) L34}, [\href{https://arxiv.org/abs/1510.07913}{{\ttfamily
  1510.07913}}].

\bibitem{dominguez11b}
A.~{Dom{\'{\i}}nguez}, M.~A. {S{\'a}nchez-Conde} and F.~{Prada},
  \emph{{Axion-like particle imprint in cosmological very-high-energy
  sources}}, \href{http://dx.doi.org/10.1088/1475-7516/2011/11/020}{\emph{JCAP}
  {\bfseries 11} (Nov., 2011) 020},
  [\href{https://arxiv.org/abs/1106.1860}{{\ttfamily 1106.1860}}].

\bibitem{dominguez13b}
A.~{Dom{\'{\i}}nguez} and F.~{Prada}, \emph{{Measurement of the Expansion Rate
  of the Universe from {$\gamma$}-Ray Attenuation}},
  \href{http://dx.doi.org/10.1088/2041-8205/771/2/L34}{\emph{APJl} {\bfseries
  771} (July, 2013) L34}, [\href{https://arxiv.org/abs/1305.2163}{{\ttfamily
  1305.2163}}].

\bibitem{CTA_LIV}
J.~{Ellis} and N.~E. {Mavromatos}, \emph{{Probes of Lorentz violation}},
  \href{http://dx.doi.org/10.1016/j.astropartphys.2012.05.004}{\emph{APP}
  {\bfseries 43} (Mar., 2013) 50--55},
  [\href{https://arxiv.org/abs/1111.1178}{{\ttfamily 1111.1178}}].

\bibitem{B0218}
M.~L. {Ahnen} et~al., \emph{{Detection of very high energy gamma-ray emission
  from the gravitationally lensed blazar QSO B0218+357 with the MAGIC
  telescopes}}, \href{http://dx.doi.org/10.1051/0004-6361/201629461}{\emph{APP}
  {\bfseries 595} (Nov., 2016) A98},
  [\href{https://arxiv.org/abs/1609.01095}{{\ttfamily 1609.01095}}].

\bibitem{PKS1441}
M.~L. {Ahnen} and et~al., \emph{{Very High Energy {$\gamma$}-Rays from the
  Universe's Middle Age: Detection of the z = 0.940 Blazar PKS 1441+25 with
  MAGIC}}, \href{http://dx.doi.org/10.1088/2041-8205/815/2/L23}{\emph{APJl}
  {\bfseries 815} (Dec., 2015) L23},
  [\href{https://arxiv.org/abs/1512.04435}{{\ttfamily 1512.04435}}].

\bibitem{1441_veritas}
A.~U. {Abeysekara}, S.~{Archambault}, A.~{Archer}, T.~{Aune}, A.~{Barnacka},
  W.~{Benbow} et~al., \emph{{Gamma-Rays from the Quasar PKS 1441+25: Story of
  an Escape}}, \href{http://dx.doi.org/10.1088/2041-8205/815/2/L22}{\emph{ApJL}
  {\bfseries 815} (Dec., 2015) L22},
  [\href{https://arxiv.org/abs/1512.04434}{{\ttfamily 1512.04434}}].

\bibitem{redshifts}
P.~{Goldoni}, others and E.~L.~f.~t. {CTA consortium}, \emph{{Redshift
  measurement of Fermi Blazars for the Cherenkov Telescope Array}},
  {\emph{ArXiv e-prints} (Aug., 2015) },
  [\href{https://arxiv.org/abs/1508.06059}{{\ttfamily 1508.06059}}].

\end{thebibliography}\endgroup

\end{document}